# Demonstrating that chlorine dioxide is a size-selective antimicrobial agent and high purity $ClO_2$ can be used as a local antiseptic

Zoltán Noszticzius, Maria Wittmann✉, Kristóf Kály-Kullai
> Group of Chemical Physics, Department of Physics, Budapest University of Technology and Economics, Budapest, Hungary
> ✉ wittmann@eik.bme.hu

Zoltán Beregvári
> Jósa András Hospital, Nyíregyháza, Hungary

István Kiss
> St. Imre Hospital, Budapest, Hungary

László Rosivall
> Semmelweis University, Budapest, Hungary

János Szegedi
> Jósa András Hospital, Nyíregyháza, Hungary

**Abstract**

Background / Aims

$ClO_2$, the so-called "ideal biocide", could also be applied as an antiseptic if it was understood why the solution's rapid killing of microbes does not cause any harm to humans or to animals. Our aim was to study both theoretically and experimentally its reaction-diffusion mechanism to find the source of that selectivity.

Methods

$ClO_2$ permeation measurements through protein membranes were performed and the time delay of $ClO_2$ transport due to reaction and diffusion was determined. To calculate $ClO_2$ penetration depths and estimate bacterial killing times, approximate solutions of the reaction-diffusion equation were derived. Additionally, as a preliminary test, three patients with infected wounds were treated with a 300 ppm high purity $ClO_2$ solution and the healing process was documented.

Results

The rate law of the reaction-diffusion model predicts that the killing time is proportional to the square of the characteristic size (e.g. diameter) of a body, thus, small ones will be killed extremely fast. For example, the killing time for a bacterium is on the order of milliseconds in a 300 ppm $ClO_2$ solution. Thus, the few minutes of contact time (owing to the volatility of $ClO_2$) is quite enough to kill all bacteria, but short enough to keep $ClO_2$ penetration into the living tissues safely below 0.1 mm, minimizing cytotoxic effects. Pictures of successful wound healings confirm these considerations. Various properties of $ClO_2$, advantageous for an antiseptic, are also discussed. Most importantly, bacteria are not able to develop resistance against $ClO_2$ as it reacts with biological thiols which play a vital role in all living organisms.

Conclusion

Selectivity of $ClO_2$ between humans and bacteria is based not on their different biochemistry, but on their different size. Preliminary clinical results encourage further research with this promising local antiseptic.

**Funding:** This work was supported by OTKA Grant 77908.

**Competing interests:** Three of the authors, namely Zoltán Noszticzius, Maria Wittmann and Kristóf Kály-Kullai declare competing financial interest as they are co-inventors of the European patent 2069232 "Permeation method and apparatus for preparing fluids containing high purity chlorine dioxide", see also reference 13. In addition Zoltán Noszticzius is also a founder of the Solumium Ltd. The other four authors have no competing financial interest.

## Introduction

The emergence and dissemination of new antibiotic-resistant bacterial strains caused by an overuse of antibiotics [1] is a global public-health concern. Methicillin Resistant Staphylococcus aureus (MRSA) [1], [2] and Carbapenem- or Extreme Drug-Resistant Acinetobacter baumannii [3], [4] are only two well known examples for such bacteria attracting world wide attention. Moreover, while the number of antibiotic resistant infections is on the rise, the number of new antibiotics is declining [1], [2]. As a result of such a dangerous situation, searches for new antimicrobial agents, as well as strategies including a switch from antibiotic to antiseptic therapies, whenever that is feasible, have been initiated.

When treating local infections of wounds, ulcers or an infected mucous membrane, the application of antiseptics instead of antibiotics is a reasonable alternative especially because bacteria are less able to develop resistance against them [5]. Presently the majority of the antiseptics used for wounds [6] are organic compounds. The most frequently applied ones [6] are chlorhexidine (chlorhexidine digluconate), octenidine (octanidine dihydrochloride), polyhexanide (polyhexametylene biguanide) and triclosan (5-chlorine-2-(2,4-dichlorphenoxy)-phenol). Notable exceptions are PVP-iodine (poly(vinylpirrolidone)-iodine complex) [6] where the active ingredient is iodine, and silver [7], both being inorganic compounds.

There are some other, less used, inorganic antiseptics such as aqueous sodium hypochlorite (NaOCl), or hydrogen peroxide ($H_2O_2$) solutions, or ozone ($O_3$) gas which have some applications in dentistry [8]. These compounds, however, are mainly used as disinfectants because they can be toxic even in low concentrations, a property seriously limiting their antiseptic applications. NaOCl, for example, one of the most commonly used components of irrigating solutions in endodontic practice, can cause poisoning and extensive tissue destruction if it is injected (inadvertently) into periapical tissues in the course of endodontic therapy [9]. $H_2O_2$ is also a double edged sword against bacteria as it also hurts living tissue [10]. Moreover, many bacteria are able to resist $H_2O_2$ as their catalase enzyme is able to decompose $H_2O_2$ rapidly [11]. Thus, beside toxicity, resistance can be also a problem even with the use of inorganic disinfectants [5]. It would be therefore reasonable to choose an antiseptic which would be free of such problems. We believe that in this respect chlorine dioxide ($ClO_2$) may be the right choice, moreover $ClO_2$ has other characteristic features favourable for antiseptic applications.

In the last twenty or more years chlorine dioxide emerged as a new and popular inorganic disinfectant. It is often referred to as „the ideal biocide" [12] because of its advantageous properties. In spite of that, as far as we know, $ClO_2$ solutions are not frequently used as antiseptic. This is because the available $ClO_2$ solutions were more or less contaminated with other chemicals applied in its synthesis and that contamination formed a major obstacle in medical applications like treating infected wounds, for example. Since 2006, however, with the help of an invention [13], it is relatively easy to produce high purity aqueous $ClO_2$ solutions. These solutions are already commercially available [14] and have been successfully used in dentistry [15] since 2008. Thus, it seems reasonable to ask the question whether the „ideal biocide" in its pure form can also be an „ideal local antiseptic" at the same time.

Such an ideal local antiseptic should satisfy many criteria. First of all, it should be safe: it should act only locally to avoid the danger of systemic poisoning and should not inflict cytotoxic effects even in the disinfected area. In this respect, it is one of the main aims of the present work to find a reasonable answer for the following intriguing question: how is it possible that contacting or even drinking $ClO_2$ solution is practically harmless for animals [16] and human beings [17], while the same aqueous solution can be a very effective and a rapid killer for bacteria, fungi, and viruses? What is the basis of this unexpected selectivity?

The answer suggested in the Results section is the following: the selectivity between humans or animals and microbes is based not on their different biochemistry, but on their different size. Denominating $ClO_2$ in the title as a „size selective" antimicrobial agent aims to emphasize this new type of selectivity. To reach that conclusion, $ClO_2$ transport was studied experimentally via protein membranes. The results of these experiments were evaluated applying a reaction-diffusion model for the $ClO_2$ transport in a reactive medium to obtain the diffusion coefficient of $ClO_2$, and the concentration of reactive groups in a protein medium. Based on these parameters the killing time, the time needed to flood a bacterium completely with $ClO_2$, can be calculated. (Details of the reaction-diffusion model and the derivation of formulae estimating the killing time are given in the Supplementary material.) It was found that the characteristic time necessary to kill a microbe is only a few milliseconds. As $ClO_2$ is a rather volatile compound its contact time (its staying on the treated surface) is limited to a few minutes. While this stay is safely long enough (being at least 3 orders of magnitude longer than the killing time) to inactivate all bacteria on the surface of the organism, it is too short for $ClO_2$ to penetrate deeper than few tenths of a millimetre; thus, it cannot cause any real harm to an organism which is much larger than a bacterium.

To show that these ideas can be applied in medical practice, we present some preliminary results of successful treatments of non-healing wounds with a high purity $ClO_2$ solution.

In the Discussion part, it is shown that $ClO_2$ can meet the safety and effectiveness requirements for a local antiseptic. Next, the chemical mechanism of the antiseptic action of $ClO_2$ is discussed and compared with that of hypochlorous and hypoiodous acids (HOCl and HOI) which are „natural" antiseptics. These hypohalous acids are used by neutrophil granulocytes, the most abundant type of white blood cells in mammals, to kill bacteria after phagocytosis. Both hypohalous acids and also $ClO_2$ attack sulfhydryl groups which play an essential role in the life processes of all living systems, e.g. in ATP synthesis [18]. That explains why bacteria were not able to develop resistance against HOCl during eons of evolution and why the emergence of $ClO_2$ resistant bacterial strains cannot be expected either. Besides this similarity, however, there are also important dissimilarities among these reagents, e.g. $ClO_2$ is more selective than HOCl. Last of all, circulation in multicellular organisms can provide some additional protection to these organisms against $ClO_2$.

# Methods

**Physico-chemical methods**
**Measurement of ClO$_2$ permeation through protein membranes**
The rate of ClO$_2$ transport was measured with the apparatus shown in Fig. 1 through two kinds of protein membranes: gelatin and pig bladder membranes, respectively. Choosing a membrane geometry for the experiments is advantageous because then the problem is „one dimensional", the concentration is a function of only one spatial coordinate *x*, which is perpendicular to the membrane, and the concentration distribution can be given as *c=c(x,t)*.

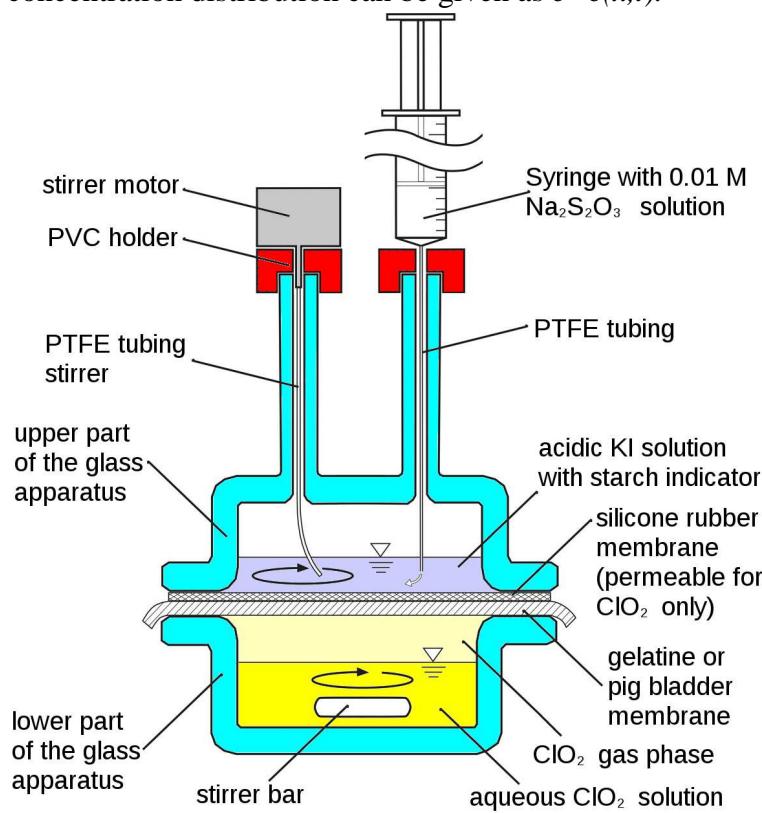

**Figure 1.** Apparatus to measure ClO$_2$ transport through gelatine or pig bladder membranes.
The two glass parts of the apparatus are held together by a pair of extension clamps (not shown in the Figure) which are fixed to a support stand by clamp holders. The active cross-section of the membranes is 28 cm$^2$. See text for the working principle.

As Fig. 1 shows, the membrane is in a horizontal position and the transport of ClO$_2$ takes place across the membrane bounded by two horizontal planes we denote by *x=0* and *x=d* in our calculations, where *d* is the thickness of the membrane.
Constant ClO$_2$ concentrations are maintained at both boundaries of the membrane, i.e. we have constant boundary conditions: $c(0,t)=c_0$ and $c(d,t)=0$, respectively. There is no ClO$_2$ in the membrane at the start of the experiment, so the initial condition: $c(0<x\leq d,0)=0$ (see Supplemental Figure S1).

While the lower face of the protein membrane is not in direct contact with the liquid phase, such direct contact would not make any difference regarding the $ClO_2$ transport. This is because the chemical potential of $ClO_2$ in the liquid and the vapour phase is the same due to the equilibrium between the liquid and the vapour phase established by continuous stirring.

Above the protein membrane there is a silicone rubber membrane in order to block the transport of any other chemicals except $ClO_2$. Silicone rubber is highly permeable for chlorine dioxide, but it is practically impermeable for other reagents [13]. This way the $ClO_2$ transport across the test membrane can be measured selectively.

Both protein membranes had a thickness of 0.5 mm and a diameter of 10 cm. The diameter of the active area in the apparatus was 6 cm resulting in an active area of 28 $cm^2$. The volume of the aqueous $ClO_2$ solution was 40 ml and its $ClO_2$ concentration was around 1000 ppm. (The exact value is given at each experiment.)

After crossing the membranes, $ClO_2$ enters the upper aqueous solution which is made by mixing 10 ml of water, 2 ml of 1 M sulphuric acid, 1 ml of 1 M KI, and 0.5 ml of 0.01 M $Na_2S_2O_3$ and as an indicator, two drops of 5 % starch solution is also added. When $ClO_2$ enters the upper solution, it oxidizes iodide to iodine, which, in turn, is reduced back to iodide again by $Na_2S_2O_3$ as long as thiosulphate is in excess. However, when all thiosulphate is consumed, the intense blue-black colour of the starch-triiodide complex appears suddenly. The time $t$ when the whole solution becomes homogeneously black (the time of the „black burst") was recorded and another 0.5 ml of $Na_2S_2O_3$ solution was added with the help of the syringe shown in the Figure. Addition of the thiosulphate eliminated the blue-black colour immediately but, after a certain period, when enough new $ClO_2$ was transported across the membrane, it reappeared again. Then the cycle was repeated starting with the injection of a new 0.5 ml portion of the $Na_2S_2O_3$ solution. The results of the measurements were depicted in a $V=V(t)$ diagram where $t$ is the time of the n-th dark burst and $V = n \times 0.5\ ml$ that is the total volume of the thiosulphate solution added before the $n$-th breakthrough.

The experiments were performed at laboratory temperature 24 ± 2 °C.

**Preparation of the gelatin membrane**

To prepare a mechanically strong membrane, it was reinforced by filter paper and the gelatin was cross-linked with glutaraldehyde. As the cellulose in the filter paper does not react with $ClO_2$ from the point of our experiments, it is an inert material.

10 ml of 10 % aqueous gelatin solution was mixed rapidly with 0.5 ml of 25 % glutaraldehyde solution at room temperature, and a filter paper disk (diameter: 10 cm) was soaked with the mixture. Then the disk was placed between two glass plates covered with polyethylene foils. Spacers were applied to produce a 0.5 mm thick membrane. After a 2 hour setting time the filter paper reinforced gelatin membrane was removed from the form and it was placed into distilled water overnight before the measurements.

**Preparation of the pig bladder membrane**

For the experiments, membrane disks with 10 cm diameters were cut from commercially available pig bladders and they were kept in distilled water for one day at +4 °C to stabilize their water content. The pig bladder membranes are slightly asymmetric: the surface of one side is smoother than the other. To obtain reproducible results, the membrane was always fixed in the apparatus with its smoother side facing downwards.

**Methods for wound healing**
**Ethics statement**
Ethical approval for the protocol and permission of the whole program (KLO2-UCD-HU_2010) was issued by the Hungarian National Health and Medical Officer Service (ANTSZ) following the suggestion of the Scientific Committee named ETT TUKEB [19] and written informed consent was obtained from all patients.

**Treatment of chronic wounds with an aqueous $ClO_2$ solution**
The solution applied contained 300 ppm high purity $ClO_2$ dissolved in distilled water (Solumium Oral® [14]). The solution was kept in a brown glass bottle which was tightly closed with a cap when not in use. Just before the treatment the cap was replaced with a spray head. For the first step of the treatment the wound was sprayed gently with the solution. (The contact with the wound does not cause discomfort as the solution does not sting.) Next one layer of non-adherent petrolatum gauze was placed on the wound together with three layers of ordinary gauze. These layers were sprayed again with the $ClO_2$ solution until they became completely wet. The bandage was finished with another three layers of dry gauze, partly to slow down the evaporation of the active ingredient and partly to decrease the patients' uncomfortable feelings due to a wet and cold bandage. The bandage was changed every 24 hours. When removing the bandage, a gentle spraying with the solution helped to detach the old gauze.

# Results

Our results cover the following themes: First, we present and evaluate membrane transport experiments aiming to determine
i) the diffusion coefficient of $ClO_2$ $D$ in a reactive protein medium, and
ii) the concentration of reactive groups $s_0$ in that medium.

To evaluate the membrane transport experiments we applied a reaction-diffusion model for the transport of $ClO_2$ in a medium containing reactive proteins. The details of that theory and the mathematical derivation of formulas applied in this section are given in the Supplemental part. Then, based on the experimentally determined $D$ and $s_0$ we calculate $T_{KILL}$, the time needed to kill bacteria by $ClO_2$, and $p$, the penetration depth of $ClO_2$ into human tissue during a wound healing treatment. Finally, to illustrate that our theory can be applied in practice, we present preliminary results of some wound healing experiments.

$ClO_2$ permeation was measured via gelatin and pig bladder membranes. The apparatus is shown in Fig. 1 of the Methods section.

**Permeation of $ClO_2$ through an artificial gelatin membrane**
Gelatin was our first choice for a model material because we wanted to study the $ClO_2$ transport in a protein medium with a known amino acid composition. Pork skin gelatin (Fluka 48719) contains only two amino acids that can react with $ClO_2$: methionine (0.88 %) and tyrosine (0.6 %) [20].

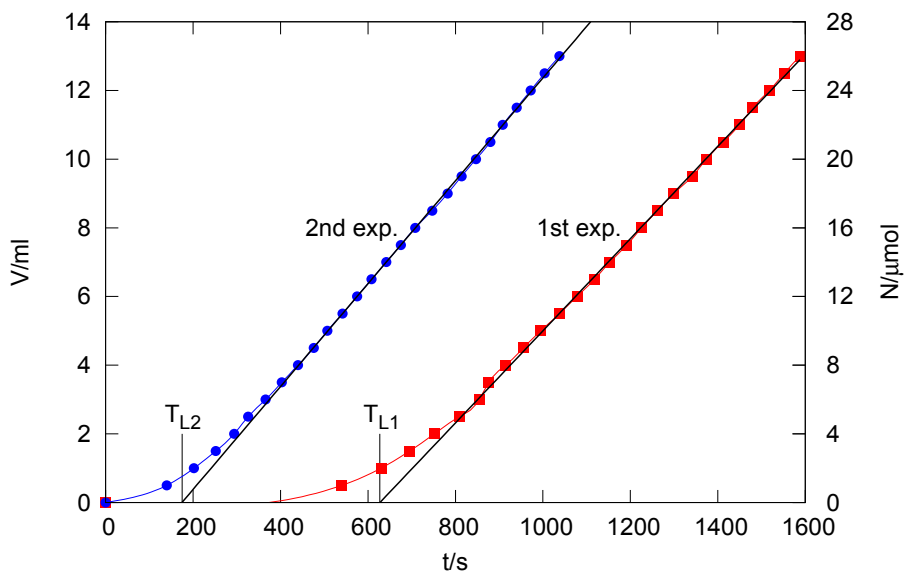

**Figure 2.** Permeation of $ClO_2$ through a gelatin membrane as a function of time $t$. Each point in the diagram represents a „black burst" (see Methods). $V$ is the cumulative volume of the 0.01 M $Na_2S_2O_3$ titrant added before the burst and $N$ is the amount of $ClO_2$ permeated until time $t$. $T_{L1} = 627$ s and $T_{L2} = 175$ s are time lags of the first and the second experiments, respectively. The concentration of $ClO_2$ source in the magnetically stirred aqueous solution was 1360 ppm (mg/kg) or 20.1 mM.

Fig. 2 shows the results of two consecutive experiments performed with the same gelatin membrane (see the two curves denoted as 1st exp. and 2nd exp.). After the first experiment, the membrane was removed from the apparatus and was kept in distilled water for 1 hour before the second experiment.

**Calculating the $ClO_2$ diffusion coefficient $D$ and the effective concentration of $ClO_2$ consuming substrates $s_0$ in gelatin**

Fig. 2. shows $N$, the cumulated amount of $ClO_2$ permeated through the membrane as a function of time. It is a common feature of both curves shown in Fig. 2 that two characteristically different dynamical regimens can be observed. In the first regimen, the amount of the permeated $ClO_2$ is very small, then, after a rapid transition period the cumulated amount of $ClO_2$ increases linearly with time. Real dynamics can be approximated with the following simplified model: zero permeation is assumed at the beginning during a waiting period but right after that a constant diffusion current appears, thus, the permeated amount increases linearly with time. To characterize such a dynamic behaviour the concept of „time lag" can be introduced: it is the time where the asymptote of the linear regimen crosses the time axis [21].

Regarding the asymptotes of the corresponding curves, the time lag in the first and in the second experiment is $T_{L1} = 627$ s and $T_{L2} = 175$ s, respectively. A logical explanation for this difference is that some $ClO_2$ is consumed inside the gelatin in the rapid reaction with methionine and tyrosine. So $ClO_2$ can break through only after it eliminates all these highly reactive amino acid residues. In the case of the second experiment, the breakthrough occurs earlier as most of these residues already reacted with $ClO_2$ during the first experiment.

If we assume that in the second experiment the reaction plays a minor role only, then in that case, the time lag is entirely due to diffusion. Roughly speaking the diffusional time lag is the time necessary to establish a steady state concentration profile inside the membrane that is to "fill up" the membrane with $ClO_2$. Based on dimensional analysis considerations (the dimension of the diffusion coefficient is (length)$^2$/(time) ) we can expect that the time lag should be proportional with the square of the thickness and inversely proportional with the diffusional coefficient. Really, the exact result [21] is that the diffusional time lag $T_{DM}$ for a membrane of thickness $d$ can be calculated as:

$$T_{DM} = \frac{1}{6} \cdot \frac{d^2}{D} \qquad (1)$$

Thus, with the assumption $T_{L2} = T_{DM} = 175$ s, $D$, the diffusion coefficient of $ClO_2$ in the gelatin membrane can be calculated knowing that $d = 0.5$ mm. The result: $D = 2.4 \times 10^{-6}$ cm$^2$s$^{-1}$.
$D$ can be determined in another way as well, from the steady state regimen. The steady state $ClO_2$ current is the slope of the curve in the linear regimen. For the 2nd experiment $J_2 = 30$ nmol/s. Then Fick's law of diffusion

$$J = A \cdot D \cdot \frac{\Delta c}{d} \qquad (2)$$

can be applied to calculate $D$. Here $A = 28.3$ cm$^2$ is the active cross-section of the membrane and $\Delta c$ is the concentration difference between the two sides of the membrane. Regarding our boundary conditions $\Delta c = c_0 = 20.1 \times 10^{-3}$ M. This way $D = 2.6 \times 10^{-6}$ cm$^2$s$^{-1}$ is obtained.
The two $D$ values, the one calculated from the time lag and the other calculated from the steady state, agree reasonably well indicating that indeed the 175 s time lag is caused mostly by diffusion and any delay due to chemical reactions is negligible in the second experiment.
On the other hand, in the first experiment, the time lag $T_{RM}$ is caused mostly by the reaction between $ClO_2$ and the reactive amino acid residues (in short "substrates") in the membrane. It is important to realize that $T_{RM}$ is not due to a slowness of the reaction kinetics (as the rate constants of the relevant $ClO_2$ – amino acid reactions are relatively high [22], [23], [24]), but it is due to the actual $ClO_2$ consumption by the reactions within the membrane delaying the breakthrough. If we assume that the rate of the chemical reaction is limited by the diffusional transport of $ClO_2$ across a zone already without reactive amino acids toward a zone of unreacted ones, then a sharp reaction front will develop on the boundary of the two zones (see Supplemental Figure S1). The front starting from one side of the membrane and driven by diffusion, propagates slowly through the membrane and $T_{RM}$ is the time when it arrives to the other side of the membrane. According to a detailed derivation in the Supplementary part, $T_{RM}$ can be given by the so-called parabolic rate law (see Supplemental equation (S12)):

$$T_{RM} = \frac{1}{2} \cdot \frac{s_0}{c_0} \cdot \frac{d^2}{D} \qquad (3)$$

where $s_0$ is the initial effective substrate concentration, i.e. the $ClO_2$ consuming capacity of the membrane in unit volume, and $c_0$ is $ClO_2$ concentration at the boundary of the membrane. Substituting the more reliable diffusion coefficient measured in the steady-state of the second experiment $D = 2.6 \times 10^{-6}$ cm$^2$s$^{-1}$ and applying the assumption that $T_{RM} = T_{L1} = 627$ s, the effective substrate concentration of the gelatin membrane $s_0$ can be calculated. The result: $s_0 = 26.2$ mM.

**Permeation of $ClO_2$ through a pig bladder membrane**

In this experiment, we studied the $ClO_2$ permeability of a pig bladder membrane which is a relatively thin (in our case it was 0.5 mm thick) but sturdy animal tissue. The same apparatus was applied as in the case of the gelatin membrane and the experimental points were depicted in Fig. 3 also with the same method.

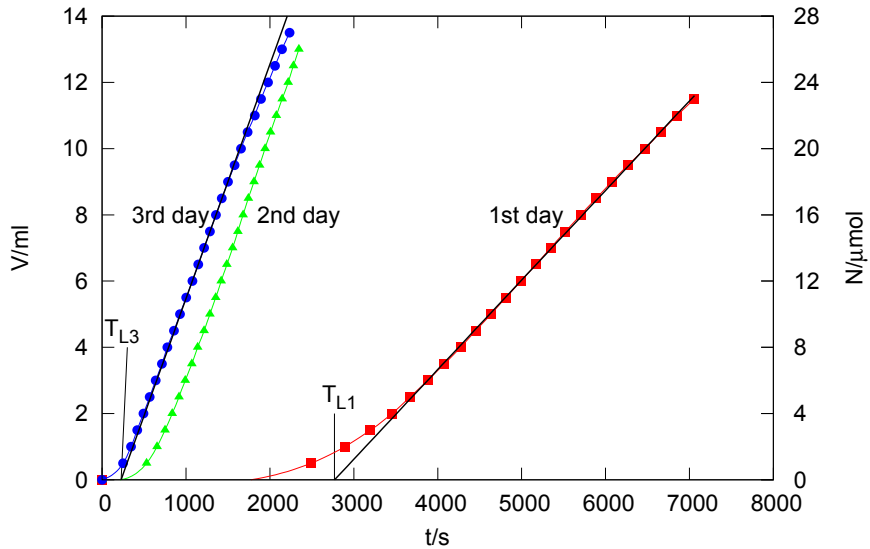

**Figure 3.** Permeation of $ClO_2$ through a pig bladder membrane as a function of time $t$.
$V$ and $N$ have the same meaning like in Fig. 2. $T_{L1} = 2770$ s, $T_{L2} = 586$ s and $T_{L3} = 226$ s are time lags of the experiments performed on the 1st, 2nd, and 3rd day, respectively. The concentration of the $ClO_2$ source was 946 ppm (14.0 mM) in these experiments.

All the three measurements (indicated as 1st day, 2nd day and 3rd day) were performed with the same pig bladder membrane but on three successive days. The membrane was kept in distilled water at +4 °C overnight between the experiments which were always started with fresh solutions.

To check the reproducibility of our measurements, we repeated the measurements with another pig bladder membrane (not shown in the Figure). While the new membrane was from a different pig bladder and its blood vessel pattern was also different, the relative deviation between the results of the two series of experiments was surprisingly small: only about 10 %. (The blood vessel structure of the membrane becomes visible as a dark network before a „black burst" because the permeability of the membrane is somewhat higher through those vessels.)

Another interesting observation was that the pig bladder membrane maintained its integrity and its mechanical strength even after the third experiment. This is because $ClO_2$ reacts selectively with certain amino acid residues of the proteins but does not destroy the peptide bonds thus the primary protein structure can survive.

**Calculating the $ClO_2$ diffusion coefficient and the effective concentration of $ClO_2$ consuming substrates in pig bladder**

Evaluation of the results was made in a similar way as in the case of the gelatin membrane. It was assumed that the time lag measured in the third experiment $T_{L3}$ =226 s is a purely diffusional time lag that is $T_{L3} \approx T_{DM}$. The diffusion coefficient of $ClO_2$ in a pig bladder membrane calculated from the above assumption is $D = 1.84 \times 10^{-6}$ cm$^2$s$^{-1}$. That value is in good agreement with the $D = 1.80 \times 10^{-6}$ cm$^2$s$^{-1}$ value calculated from the steady state current $J_3 = 14.1$ nmol/s of the 3rd experiment.

As we can see, the diffusion coefficient of $ClO_2$ in a pig bladder tissue is only 30 % smaller than in the unstructured gelatin. This supports our assumption that the cellular structure of the pig bladder tissue does not matter too much from the point of the diffusional transport of $ClO_2$ as it can penetrate through the external and internal lipid membranes of the individual cells of the tissue.

However, there is a more significant deviation between the pig bladder and the gelatin regarding $s_0$, the effective substrate concentration. Assuming that the time lag in the first experiment $T_{L1}$ = 2770 s is due to the chemical reaction, $T_{L1} \approx T_{RM}$, then from (3) we get $s_0$ = 56 mM, indicating that the concentration of the reactive components in the pig bladder tissue is about two times higher than that is in the gelatin. This is a reasonable result as the animal tissue is denser, and it contains not only methionine and tyrosine like gelatin, but also cysteine and tryptophan residues. We would like to add that in a series of measurements performed with the same membrane the steady state $ClO_2$ current in the first experiment is always smaller than in the subsequent ones, although the $ClO_2$ source is not changed. This effect is more pronounced in the case of an animal membrane (compare the slope of the 1st day experiment with that of the other days). The phenomenon can be understood if we assume that some components, which are able to react with $ClO_2$ but only slowly, can remain in the pig bladder even after the first $ClO_2$ breakthrough. As it is shown in the Supplemental equation (S40), the slow $ClO_2$ consumption of these components can explain a smaller quasi-steady state current. The fact that these components disappear from the membrane after keeping it in water overnight suggests that they are reaction products which can be leached out from the membrane or are unstable intermediates which decompose.

**Estimating the killing time for bacteria with cylindrical and spherical geometries**

We assume that a bacterium is killed when its whole volume is flooded by $ClO_2$. To calculate the killing time, if we know the shape and the size of the bacterium, we would need two more parameters, the diffusion coefficient of $ClO_2$ $D$ and the effective concentration of $ClO_2$ consuming substrates $s_0$ in the bacterial medium. In the absence of bacterial data it will be assumed that the parameters $D$ and $s_0$ in the single cell of a bacterium are close to that what we have measured above in the animal cell aggregates of the pig bladder. A further simplifying assumption is that only spherical and cylindrical bacteria are considered. Numerical results are calculated for a diameter of 1 μm, which is a characteristic length-scale for bacteria. Mathematical formulas for the killing time and the penetration depth are derived in the Supplemental part. In this section only the results of those derivations will be given together with some qualitative explanations on their meaning.

It will be assumed that the rate of the "$ClO_2$ – bacterium reaction" is also limited by the diffusion of $ClO_2$ to the fast reacting amino acid residues fixed in protein molecules like in the case of the much larger membranes and this way a sharp reaction front propagates from the cell wall toward the centre of the bacterium.

Intuitively, the killing time $T_{KILL}$ should be analogous to the time lag $T_{RM}$ in a membrane caused by a chemical reaction, because these are the times needed to flood the whole volume. We can expect, however, that the geometric factor should be different depending on the shape of the bacterium. For a cylindrical bacterium with a diameter of $d$ the killing time is

$$T_{KILL,C} = \frac{1}{16} \cdot \frac{s_0}{c_0} \cdot \frac{d^2}{D} \quad (4)$$

see supplemental equation (S18),
and for a spherical bacterium also with a diameter of $d$ it is

$$T_{KILL,S} = \frac{1}{24} \cdot \frac{s_0}{c_0} \cdot \frac{d^2}{D} \quad (5)$$

according to supplemental equation (S24). We can see that (4) and (5) are analogous to (3) but the geometric factors for a cylinder and for a sphere are much smaller than for the planar membrane indicating that in these geometries the surface from where diffusion current is starting is relatively larger compared to the volume that has to be flooded.

Substituting the pig bladder parameters $D = 1.8\times10^{-6}$ cm$^2$s$^{-1}$ and $s_0 = 56$ mM into formulas (4) and (5) together with the ClO$_2$ concentration applied in the wound healing experiments (see later) $c_0 = 4.45$ mM (Solumium Oral®, 300 ppm) and using $d = 1$ μm we obtain that the killing time for a cylindrical bacterium with a diameter of 1 μm is

$$T_{KILL,C} = 4.4 \text{ ms},$$

while the killing time for a spherical bacterium with a diameter of 1 μm is

$$T_{KILL,S} = 2.9 \text{ ms}.$$

As we can see, the killing time for a bacterium is only a few ms due to its small size. Even if $s_0$, the effective substrate concentration of a bacterium would be an order of magnitude higher than we assumed, the killing time would be still less than 0.1 s. Other approximations applied in our calculations can only overestimate the real killing time. For example, the diffusion coefficient of ClO$_2$ in the pig bladder was measured at 24 ± 2 °C. If ClO$_2$ is used to disinfect a living human tissue, the temperature is higher, which means a larger diffusion coefficient and an even shorter killing time. Another approximation is the concept of fixed substrates. Inside a bacterium mobile substrates like glutathione [25], free amino acids and various antioxidants also occur. These small molecules can diffuse by and large freely within the bacterium. Nevertheless $T_{KILL}$ would still work as a good upper estimate because the mobility of the substrate can only shorten the time needed for ClO$_2$ to reach these substrates and react with them. Furthermore, when the killing time $T_{KILL}$ is regarded as the time when the sharp front reaches the center of the sphere or the symmetry axis of the cylindrical bacterium, it will surely be overestimated, as it is not necessary to oxidize all the available substrate content of a bacterium to kill it. For example, it is enough to oxidize less than 40 % of the methionine content of E. coli to achieve a 100 % kill [26].

**Contact time and penetration depth of ClO$_2$ into human skin or wound**

When an organism is not submerged in the aqueous ClO$_2$ solution but the solution is applied on its surface only, as in the case of disinfecting wounds, the volatility of ClO$_2$ also has to be taken into account. The effective contact time is much shorter using a ClO$_2$ solution than with less or non-volatile disinfectants. According to our measurements, when a wound is covered with 3 wet and 3 dry layers of gauze more than 80 % of ClO$_2$ evaporates from the bandage within one minute due to the high volatility of ClO$_2$ and to the high specific surface of the gauze. Thus, to give an upper limit for the penetration depth into the human tissue, we will assume that the initial ClO$_2$ concentration ($c_0 = 4.45$ mM, Solumium Oral®) is maintained for 60 s, that is $T_{CON} = 60$ s, where $T_{CON}$ denotes the contact time. As a zero-th estimate, we assume again that the human tissue has the same $D$ and $s_0$ values like that of the pig bladder tissue.

Applying the parabolic rate law (see Supplemental equation (S13) where $t = T_{CON}$) the penetration depth $p$ can be estimated:

$$p = \sqrt{\frac{2c_0 D \cdot T_{CON}}{s_0}} \qquad (6)$$

$p(T_{CON} = 60\ s) = 41.5$ μm. We remark that (6) can be derived from (3) directly if we realize that for the present problem $d = p$ and $T_{RM} = T_{CON}$.

Nevertheless, the actual penetration depth into a living tissue - either its surface is a wound or an intact human skin – should be even much smaller than the above estimate. This is due to the capillary circulation which is present in living tissue but is absent from dead tissue like the pig bladder membrane used for the measurements. The serum in the blood vessels and also the extracellular fluid contain many components capable of reacting rapidly with $ClO_2$. The fluid transport of these reactive components in the blood capillaries of the dermis [27] can maintain a finite reactant concentration in that region. Then the diffusive transport of these reactants outward from the dermis into the epidermis [27] can halt an inward propagating reaction front establishing a steady state.

Moreover, in the case of intact human skin, $ClO_2$ should permeate through the stratum corneum [28] first, which is the 10–40 μm thick outermost layer of epidermis consisting of several layers of dead cells. This keratinous layer forms a barrier to protect the underlying tissue from infection, dehydration and chemicals. The diffusion coefficient of $ClO_2$ in that layer should be much lower compared to the underlying tissue.

As we can see, the penetration depth into human skin is only few tens of a micrometer even if we neglect circulation. Such shallow penetration cannot really harm human tissues. On the other hand, this short contact time is still several orders of magnitude larger than the killing time, $T_{CON} \gg T_{KILL}$, which is the necessary criterion of a successful disinfection.

**Therapeutic window**

The above formulas and calculations indicate that disinfection of living tissues with aqueous $ClO_2$ solutions has a very wide therapeutic window: while surprisingly low concentrations and short contact times are able to kill bacteria, much higher concentrations and residence times are still safe to use.

There is one notable exception: inhaling high concentration $ClO_2$ gases for an extended time can be dangerous for human health because the alveolar membrane is extremely thin (a mere 1-2 microns and in some places even below 1 micron). The effect of $ClO_2$ in these membranes is somewhat counterbalanced, however, by the intense blood circulation there.

**Treatment of non-healing wounds with a dilute $ClO_2$ solution**

A medical research program was approved in 2011 in two hospitals in Hungary to test how safe and effective $ClO_2$ is as a local antiseptic. The aim of this program among others is to treat otherwise non-healing wounds with a dilute high purity chlorine dioxide solution (300 ppm $ClO_2$, Solumium Oral®). Until now more than 50 patients with chronic wounds were treated successfully. The following three Figures are preliminary results taken from a longer and more detailed report, the preparation of which is in progress. A bacterial spectrum belonging to each Figure is given in Supplementary Table S3.

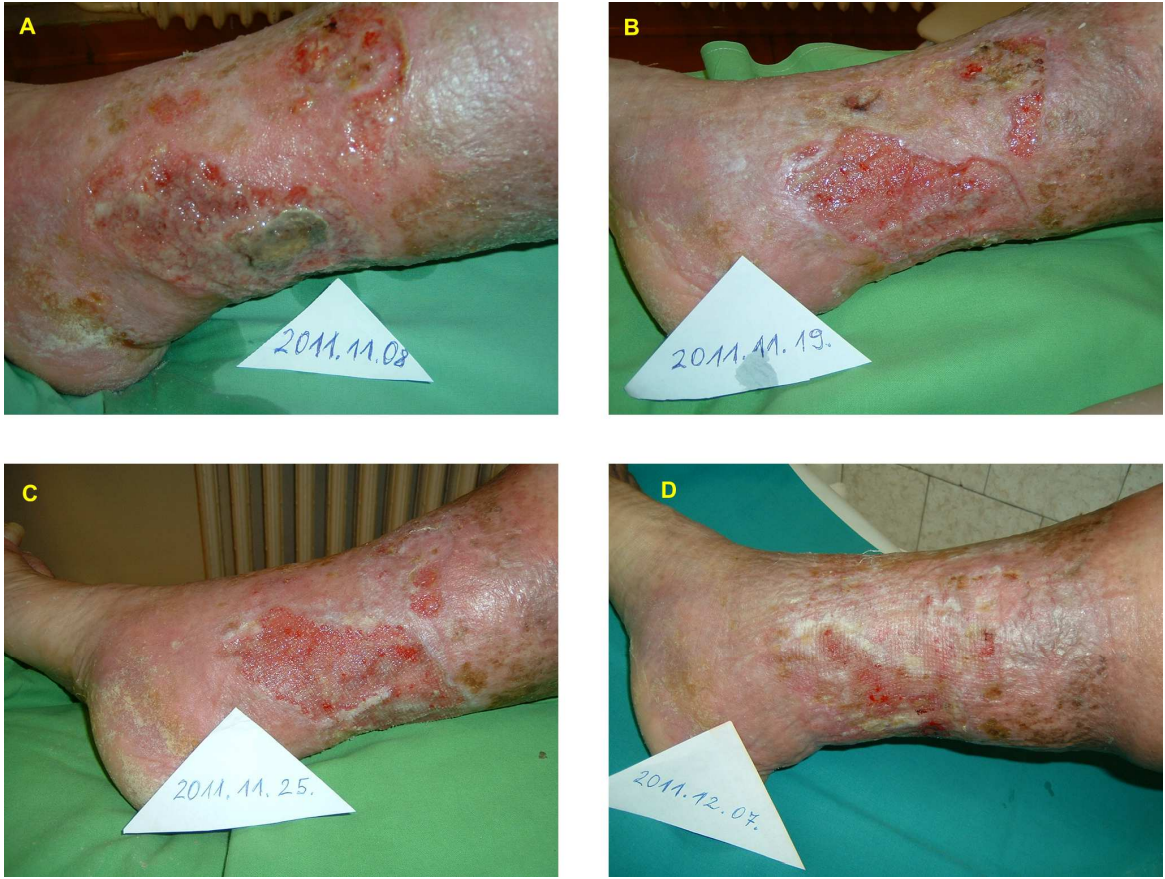

**Figure 4.** Patient I.
65 years old female patient who got deep vein thrombosis some years ago. She was treated with recurrent ulcer several times, and in this state she was admitted to the hospital.
a) Nov 8 2011: The state before Solumium solution treatment was started.
b) Nov 19 2011: Status during treatment, the clearance and the granulation of the wound is definitely visible. The necrotic biofilm was gone. From this point the wound dressings were completed also with flexible swaddling.
c) Nov 25 2011: Beside the wound granulation also the epithelialisation in the edge region has started.
d) Dec 7 2011: Epithelialisation progressed and the ulcer was healed almost completely.

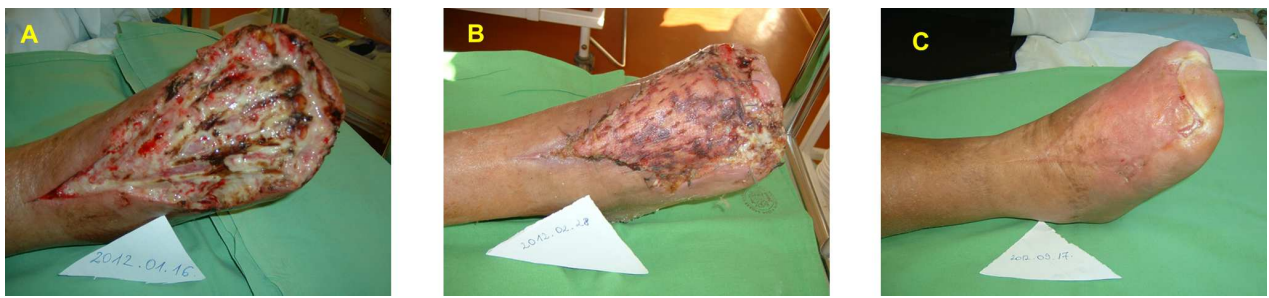

**Figure 5.** Patient II.

60 year old woman having severe diabetes for years, treated with insulin. She was admitted to hospital and operated on festering and necrosis of the toes that developed on the basis of the foot neuropathy and angiopathy.

a) Jan. 16 2012: The image shows the status after the removal of the toes and the resection of dead skin and subcutaneous tissue. At this point Solumium solution treatment was started

b) Feb. 28 2012: After granulation of the wound base it was covered with a half-thick lobe of skin and Solumium solution treatment was continued.

c) Sept. 17 2012: The healed state. (The actual wound healing had been completed much earlier. The picture was taken later when the patient returned to the hospital to check her state.)

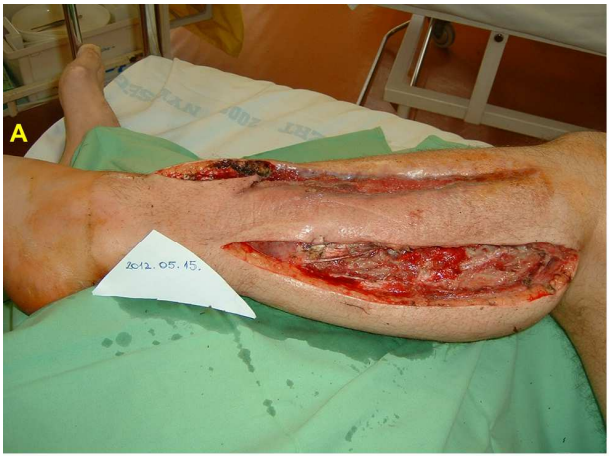
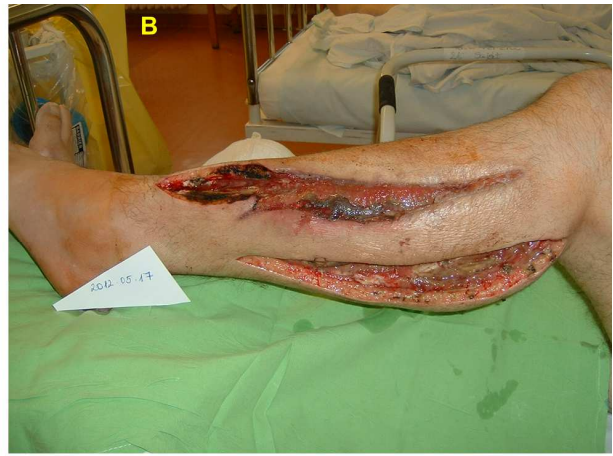
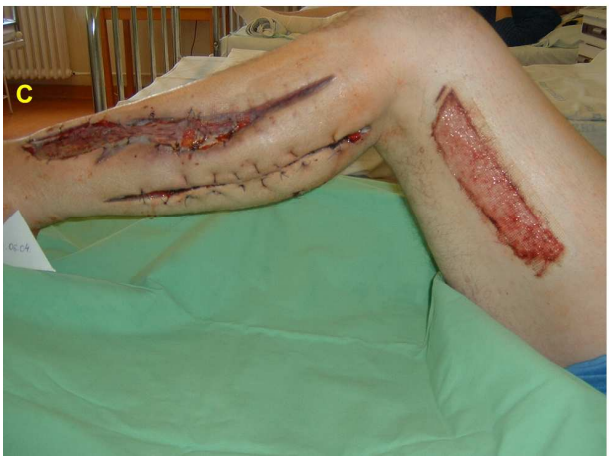
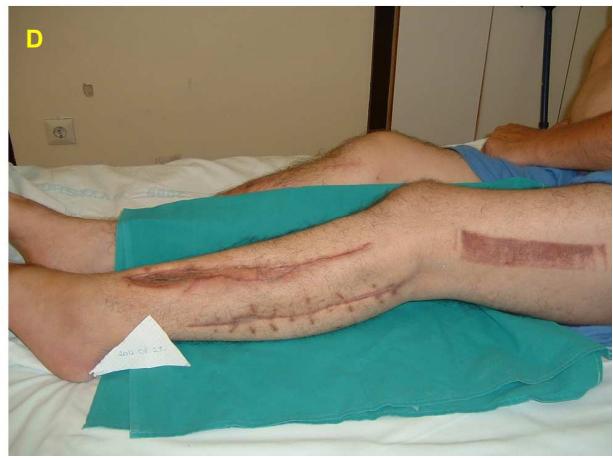

**Figure 6.** Patient III.

52 years old man with tablet-treated diabetes. He suffered a motor accident in April 2012. An extensive hematoma was formed under the leg skin that suppurated and also the plate of the connective tissue covering the muscles necrosed.

a) May 15 2012: The picture shows the status after surgical exploration in which the necrosed tissue was excised. The wounds were kept open and Solumium treatment was started.

b) May 17 2012: Condition during treatment, the wound base started granulating

c) June 16 2012: After the clearance of the lower wound it could be closed with direct sutures. The upper wound which also cleared up and granulated was covered with a half-thick lobe of skin taken from the thigh. Solumium treatment was continued

d) August 27 2012: The healed state. (The actual wound healing had been completed much earlier. The picture was taken later when the patient returned to the hospital to check his state.)

The preliminary results prove that aqueous $ClO_2$ solutions can be applied to heal chronic wounds. To compare the performance of $ClO_2$ with other antiseptics requires further study, which is planned. It is encouraging, however, that all treated patients were healed and $ClO_2$ caused no inflammatory response in any patients.

# Discussion

In this section first we discuss whether $ClO_2$ should be regarded as an "exotic" antiseptic only or it has the promise to become a commonly used antiseptic to treat local infections. To this end safety and effectiveness requirements for a local antiseptic are collected to check how $ClO_2$ can meet these requirements compared to other antiseptics.

Next a biochemical action mechanism, explaining the antiseptic effect of $ClO_2$ is discussed, which is partly analogous to that of hypochlorous and hypoiodous acids. These "natural" antiseptics also react, among others, with sulfhydryl groups like $ClO_2$ but their reaction products can be different. The importance of that difference and the protective role of SH groups and of the circulatory system, existing in a multicellular organism only, is also discussed.

**Safety and effectiveness requirements for a local antiseptic**

A local antiseptic should meet the following requirements to be considered
as safe:
i)      it should act only locally to avoid systemic poisoning, and
ii)     it should not prevent or delay the process of healing, i.e. it should not be cytotoxic.
and as effective:
iii)    it should be effectual in relatively low concentrations, and even in biofilms (biofilms are medically important, accounting for over 80 percent of microbial infections in the body [29]) as well, and
iv)    microbes should not be able to develop resistance against it (a problem related to the biochemical mechanism of action).

As it was shown in the Results section $ClO_2$ as a size selective antiseptic, meets requirements i) and ii). Thus only criteria iii) and iv) are discussed here.

**Comparing the biocidal activity of $ClO_2$ to that of other antiseptics (criterion iii)**

In free aqueous solutions, the strongest chemical disinfectant is ozone. In biofilms, however, the performance of ozone is rather poor. In addition, ozone is toxic and decomposes in aqueous solutions rapidly. (Its half life is only 15 min at 25 °C at pH 7.) All of these disadvantageous properties of ozone prevent its use as an antiseptic in most applications.

The second strongest disinfectant after ozone is chlorine dioxide. Tanner [30] made a comparative testing of eleven disinfectants on three test organisms (including two bacteria: Staphylococcus aureus and Pseudomonas aeruginosa and one yeast: Saccharomyces cerevisiae).

He found that the disinfectant containing $ClO_2$ had the highest biocidal activity on a mg/l basis against the test organisms. Beside antibacterial and antifungal properties, $ClO_2$ also shows strong antiviral activity, about ten times higher than that of sodium hypochlorite [31]. And it inactivates practically all microbes including algae and animal planktons [32] and protozoans [33]. Moreover $ClO_2$ can remove biofilms swiftly [12] because it is highly soluble in water and unlike ozone it does not react with the extracellular polysaccharides of the biofilm. This way $ClO_2$ can penetrate into biofilms rapidly to reach and kill the microbes living within the film.

**Impossibility of bacterial resistance against $ClO_2$ (criterion iv)**

$ClO_2$ is a strong, but a rather selective oxidizer. Unlike other oxidants it does not react (or reacts extremely slowly) with most organic compounds of a living tissue. $ClO_2$ reacts rather fast, however, with cysteine [22] and methionine [34] (two sulphur containing amino acids), with tyrosine [23] and tryptophan [24] (two aromatic amino acids) and with two inorganic ions: $Fe^{2+}$ and $Mn^{2+}$. It is generally assumed that the antimicrobial effect of $ClO_2$ is due mostly to its reactions with the previously mentioned four amino acids and their residues in proteins and peptides. In the peptide group it is important to mention glutathione – a small tripeptide containing cysteine - which is a major antioxidant in cells, with an intracellular concentration of 0.1-10 mM [35].

Margerum's group [22], [23], [24] reported the following second order rate constants at pH 7 and 25 °C: cysteine $1\times10^7$ $M^{-1}s^{-1}$ >> tyrosine $1.8\times10^5$ $M^{-1}s^{-1}$ > tryptophan $3.4\times10^4$ $M^{-1}s^{-1}$. As can be seen, cysteine is the far most reactive amino acid because of its thiol group. As the above mentioned four amino acids and especially cysteine and biological thiols play a crucial role in all living systems, including microbes, it is impossible for any microbe to develop a resistance against chlorine dioxide.

As an important analogy we can mention that bacteria have never been able to become resistant against hypochlorous acid (HOCl) either, which is an important natural antiseptic used by neutrophils for millions of years. Neutrophils, a type of white blood cells, are phagocytes which kill the engulfed microbes by applying various hydrolytic enzymes and hypohalogeneous acids, chiefly HOCl [36], [37]. On that basis Robson and co-workers applied HOCl as a kind of „natural" wound care agent [38], [39]. Thus, it is reasonable to compare the action mechanisms and other properties of $ClO_2$ and HOCl as antiseptic agents.

**Comparison of $ClO_2$ and HOCl as possible antiseptic agents**

HOCl, like $ClO_2$, reacts rapidly with the sulphur containing amino acid residues of methionine and cysteine, the second order rate constant (at pH 7.4 and 22 °C) being $3.8\times10^7$ $M^{-1}s^{-1}$ and $3.0\times10^7$ $M^{-1}s^{-1}$, respectively, and also reacts with tryptophan ($1.1\times10^4$ $M^{-1}s^{-1}$) and tyrosine (44 $M^{-1}s^{-1}$) [40]. However, unlike $ClO_2$, HOCl reacts rapidly with many other amino acid residues and even with peptide bonds [40], and many other compounds such as carbohydrates, lipids, nucleobases, and amines [41].

As we can see the important similarity is the fast reaction of both HOCl and $ClO_2$ with the SH group of cysteine. This is important because it is assumed that abolition of ATP synthesis and killing bacteria by HOCl is due to its reaction with sulfhydryl groups [18]. It is a logical assumption that $ClO_2$ can also stop the ATP synthesis as it reacts with the very same SH groups like HOCl.

At the same time, however, there are important dissimilarities between HOCl and $ClO_2$:

i) HOCl is much less specific and reacts rapidly with numerous other substrates. Thus killing bacteria with HOCl requires more reagent than with $ClO_2$.

ii) While $ClO_2$ evaporates rapidly from its aqueous solution and can reach and kill bacteria even through a gas phase, e.g. through an air bubble blocking a dental root canal [42], evaporation of HOCl is not significant. Thus HOCl stays at the disinfected area for a long time even after killing all bacteria which can cause inflammation there [43].

iii) HOCl is a more drastic reagent and causes irreversible damage. For example $ClO_2$ oxidizes glutathione (GSH) mainly to glutathione disulfide (GSSG) [22] which can be reduced back to GSH easily in a natural way in the body. On the other hand, HOCl can attack disulfide bonds and oxidizes GSH mostly to glutathione sulfonamide (GSA) [44] causing an irreversible loss of the cellular GSH.

**Sulfhydryl groups and circulation can protect multicellular organisms from $ClO_2$ inflicted irreversible damage**

As it was mentioned, the $ClO_2$ –SH group reaction has the highest rate constant among the $ClO_2$ – amino acid reactions. (Cysteine or GSH [22] reacts about 50 times faster than the runner up tyrosine.) Consequently, as long as some SH groups are present (mostly in the form of GSH), these groups react with $ClO_2$ rapidly protecting other amino acid residues from oxidative damage. Moreover the oxidation of SH groups to disulfide bonds can be reversed. An interesting example was presented by Müller and Kramer [45], [46]. They found that the cytotoxic effect of povidone-iodine after a 30 min contact with murine fibroblast was only temporal: after a 24 hour culture without the antiseptic an unexpected revitalization of the fibroblasts was observed [45]. According to Winterbourn and co-workers [47], HOI (the reactive hydrolysis product of iodine) also oxidises GSH to GSSG but not to GSA. That parallelism between the reversible HOI-GSH and the $ClO_2$–GSH reactions raises the question whether an analogous revitalization might be also possible in the case of $ClO_2$. This question is all the more justified since in some animal experiments [16] rats were drinking water containing 200 ppm $ClO_2$ for 90 days but without developing any gastrointestinal problems. In those experiments all $ClO_2$ must have reacted with the animal tissues as it cannot evaporate from the stomach of the rats. To interpret that result it is reasonable to assume that SH groups transported by the circulation system of the rodent protected the epithelial cells in its gastrointestinal tract from an irreversible oxidation by $ClO_2$.

Above a certain limit, however, when a too high percentage of the protective SH groups is already oxidized, $ClO_2$ would inflict irreversible changes to the higher order protein structures by oxidizing the tyrosine and tryptophan residues [48]. That would certainly happen with the bacteria on the surface of an infected tissue as their GSH supply [26] can be rapidly exhausted by $ClO_2$. Mammalian cells below the surface, however, might survive being supported by the circulation which transports protective sulfhydryl and other reductive compounds to the cells, continuously repairing or even revitalizing them.

Thus beside their size there is another important difference between single cell and more complex multicellular organisms: it is the circulation which can help the cells of a multicellular organism to survive while that type of help is not available for a bacterium.

## Conclusion

Chlorine dioxide is a size selective antimicrobial agent which can kill micron sized organisms rapidly but cannot make real harm to much larger organisms like animals or humans as it is not able to penetrate deeply into their living tissues. Moreover the circulation of multicellular organisms can provide an additional protection to these organisms against $ClO_2$.

The predicted wide therapeutic $ClO_2$ concentration window in antiseptic applications was tested in wound healing experiments with promising results. These results are preliminary ones, however, and a detailed experimental study to compare high purity aqueous $ClO_2$ solutions with other antiseptics is needed. It is an aim of the present work to initiate such studies hoping that $ClO_2$ could be applied to treat various local infections, especially where bacterial resistance is a problem.

**Acknowledgments**

The authors thank Ms. Marianna Megyesi for performing the physico-chemical experiments and Dr. György Berencsi, Dr. Mihály Kádár, Dr. Márta Milassin, Prof. Ferenc Rozgonyi and Dr. Gusztáv Várnai for helpful discussions. We also thank Ms. Mary-Beth Sinamon for proofreading our MS as a native English speaker.

**Author Contributions**

Z. N. conceived and designed the physico-chemical experiments and wrote the manuscript, M. W. and K. K. contributed to the reaction-diffusion model and mathematical derivations and performed data analysis, Z. B. treated the chronic wounds shown in the paper and made the photos to document the process of healings, I. K., L. R. and J. Sz. organized the medical research, performed preliminary tests and wrote the protocols for the wound healing experiments. All authors contributed to discussions and reviewed the manuscript.

# Supplementary Information

**A reaction-diffusion (RD) model for the transport of ClO₂
in a medium containing reactive proteins**

A general RD equation for ClO₂
The following partial differential equation (usually called reaction-diffusion equation[1]) holds for the local ClO₂ concentration $c$ ($c$ is a function of the time $t$ and of the space coordinates) when ClO₂ diffuses through a medium containing various components which can react with it:

$$\frac{\partial c}{\partial t} = -\sum_{i=1}^{N} R_i + D\nabla^2 c . \qquad (S1)$$

In equation (S1) $\frac{\partial c}{\partial t}$ is the time derivative of the local ClO₂ concentration, $R_i$ is the rate of the ClO₂ consumption due to the $i$-th reaction at the same location, $N$ is the number of the various ClO₂ consuming reactions, $D$ is the diffusion coefficient of ClO₂ in the medium, and $\nabla^2 c$ is the Laplacian of $c$, which – applying a three dimensional Descartes coordinate system with spatial coordinates $x$, $y$, and $z$ – can be written in the following form:

$$\nabla^2 c = \frac{\partial^2 c}{\partial x^2} + \frac{\partial^2 c}{\partial y^2} + \frac{\partial^2 c}{\partial z^2} . \qquad (S2)$$

Equation (S1) is a balance equation for ClO₂ where the two terms on the right hand side stand, successively, for the effect of the chemical reactions and of diffusional transport[1].

A simplified RD equation for ClO₂. The effective substrate concentration
As it was discussed previously there are four different amino acids and amino acid residues which can react with ClO₂ rapidly. In living tissue, however, there are even more chemical components[2] which are also able to react with ClO₂ by a slower but still measurable rate. A simple model cannot deal with all the ClO₂ reducing substrates of a complex biological system individually. To simplify the model the concept of the effective substrate concentration $s$ will be introduced, which represents the local ClO₂ reducing capacity of all the various substrates in an integrated form.

To develop a definition for $s$, let us write the stoichiometry of the $i$-th reaction (the reaction of the $i$-th substrate $S_i$ with ClO₂) in the following simplified form:

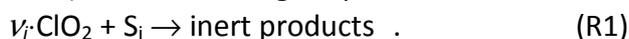
$\nu_i \cdot ClO_2 + S_i \rightarrow$ inert products . (R1)

The stoichiometric coefficient $\nu_i$ shows how many ClO₂ moles can be reduced by one mole of $S_i$. For example, when the substrate contains an SH (sulfhydryl or thiol) group as cysteine does, the stoichiometric equation around pH 7 for a fast initial reaction[3] can be written as

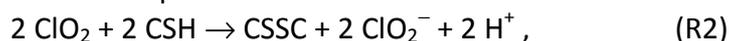
$2\ ClO_2 + 2\ CSH \rightarrow CSSC + 2\ ClO_2^- + 2\ H^+$ , (R2)

where CSH stands for cysteine and CSSC is its oxidation product, a disulfide, called cystine. (One of the products, $ClO_2^-$ (chlorite) is actually an intermediate because it can react further with cysteine, but only with a rate which is 6 orders of magnitude slower than the first step of the ClO₂/CSH reaction[3].) Thus, if we regard only the fast initial reaction then $\nu_{CSH} = 1$, because 1 mole CSH removes 1 mole ClO₂ in (R2). For tyrosine[4] and tryptophane[5], a simplified scheme would suggest $\nu_{TYR} = \nu_{TRP} = 2$. Even in these relatively simple cases of pure amino acids, however, the effect of various parallel and consecutive reactions[3,4,5] can make it rather difficult to calculate $\nu$ very precisely, not to mention when these amino acids are residues in proteins or peptides.

For the definition of $s$, however, it is enough to assume that there is such a stoichiometric coefficient for each component. Then $s$, the effective substrate concentration of the medium, can be defined as a weighted sum of the individual $s_i$ substrate concentrations, where $\nu_i$ plays the role of a „weight factor":

$$s = \sum_{i=1}^{N} \nu_i \cdot s_i \,. \qquad (S3)$$

Moreover, as a further simplification, it will be assumed that $R_i$, the rate of $ClO_2$ reduction due to the $i$-th reaction can be written as a bilinear function of $s_i$ and $c$:

$$R_i = \nu_i \cdot k_i \cdot s_i \cdot c, \qquad (S4)$$

where $k_i$ is the second order rate constant of the $i$-th reaction. This way we assume that the rate determining step of the $ClO_2$ – substrate reaction has second order kinetics. The assumption is certainly true for cystein, tyrosine and tryptophan substrates, as it was shown by Margerum's group [3,4,5]. Next, introducing an „effective rate constant" $k$ by the definition:

$$k = \frac{\sum_{i=1}^{N} \nu_i \cdot k_i \cdot s_i}{\sum_{i=1}^{N} \nu_i \cdot s_i}, \qquad (S5)$$

equation (S1) has the following simple form:

$$\frac{\partial c}{\partial t} = -k \cdot s \cdot c + D\nabla^2 c \,. \qquad (S6)$$

Simplified balance equations for fixed substrates

We will also assume all the substrates are fixed to the medium, and it is only the $ClO_2$ which is able to diffuse. This approximation is reasonable, if the RD medium is a human or an animal tissue having a cellular structure. Amino acid residues are usually parts of large protein molecules, the diffusion of which is very slow. Smaller peptides – especially glutathione – and free amino acids can diffuse but only within a cell because the outer membrane of the cell is not permeable for them. Thus, from the point of a long range transport through an animal or human tissue, even these small substrates can be regarded as fixed ones. This way, the general RD equation for a substrate is

$$\frac{\partial s_i}{\partial t} = -k_i \cdot s_i \cdot c + D_i \nabla^2 s_i \qquad (S7)$$

can be simplified to

$$\frac{\partial s_i}{\partial t} = -k_i \cdot s_i \cdot c \qquad (S8)$$

as all $D_i = 0$. If we multiply both sides of equation (S8) with $\nu_i$, and then summarize all such type of equations then we obtain the balance equation for the effective substrate concentration in the following simple form:

$$\frac{\partial s}{\partial t} = -k \cdot s \cdot c \,. \qquad (S9)$$

When the medium contains the very reactive SH groups in a significant concentration then it can be proven that (S6) and (S9) can be simplified further: the form of the equations remains the same but the effective rate constant can be approximated as

$$k \approx k_{SH} \qquad (S10)$$

where $k_{SH}$ is the rate constant of the ClO$_2$ – SH group reaction, and the effective substrate concentration is

$$s \approx s_{SH} \qquad (S11)$$

where $s_{SH}$ is the concentration of the sulfhydryl groups in the medium.

Approximate solutions of the simplified RD equations
If the simplified RD equations (S6) and (S9) are accepted as a starting point, then the logical next step is to find a solution for these equations, that is to find the functions $c=c(t,x,y,z)$ and $s=s(t,x,y,z)$ while taking into account the given initial and boundary conditions. To find exact analytical solutions for nonlinear partial differential equations is usually not possible and in this work we did not want to apply numerical solutions either. Thus, our aim here should be to find and apply approximate solutions with simple mathematical formulas which can be easily applied for the interpretation of our experimental results.

One type of approximation can be applied when the rate constant $k$ is very high, as in the case of substrates containing SH groups or tyrosine residues. In this case, a sharp reaction front propagates through the medium, and the solution of the reaction-diffusion problem can be approximated with „parabolic rate law" type equations.

The other approximation is valid for low $k$ values. In this case, the smooth concentration profiles are determined mostly by the diffusion, and modified only slightly by the reaction which can be distributed in the whole medium (no sharp front). When the medium is finite, as in the case of a membrane, an approximate steady state can be reached after some transition time.

## Quasi steady state solution of the RD equations
## when the ClO$_2$ – substrate reaction is fast

Preconditions of the parabolic rate law
The so-called parabolic rate law[6] holds for certain reaction-diffusion problems where the rate limiting step of an otherwise fast irreversible reaction (R3)

$$C + S \rightarrow P \qquad (R3)$$

between the mobile reactant C and the fixed substrate S giving the product P is not the reaction itself but the diffusion of the reactant C to reach S. In our case C is ClO$_2$ and S is the reactive side group of an amino acid. The most important reactant in this respect is the SH group of the cysteine[3].

An important player in the process is the medium M immobilizing the substrate S but permeable for C at the same time. In our case the medium M is the hydrogel of the living tissues which is permeable for ClO$_2$. Lipid membranes of the cells in the tissue do not form barriers for ClO$_2$ either, as it is very soluble in organic phases as well. The reactive amino acids, on the other hand – being mostly building blocks of various proteins – are immobilized in that hydrogel.

The parabolic rate law in one dimension for a slab of thickness $d$
The simplest geometry giving a parabolic rate law is a situation where the concentration of C is kept constant, [C] = $c_0$ at the flat boundary of a slab e.g. at its left hand side, while [C] = 0 at the right hand side of the slab. The material of the slab is a medium containing the fixed substrate S in a homogeneous initial concentration $s_0$ (see Fig. S1). The thickness of the slab is $d$.

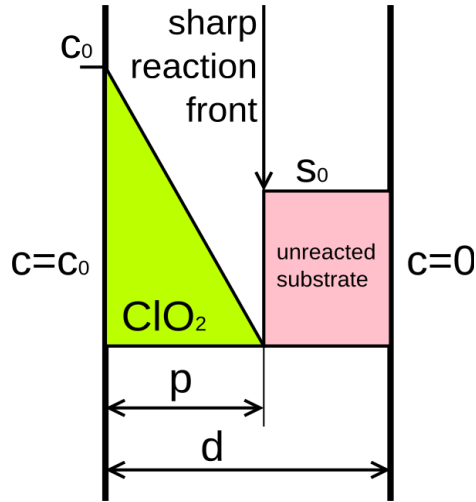

**Figure S1. Schematic ClO$_2$ and substrate concentration profiles in a hydrogel slab of thickness *d* at an intermediate time *t* ( 0 < *t* < *T* ), where *p* is the penetration depth.**

When C is ClO$_2$ and ClO$_2$ is fed from one side of the slab, a sharp reaction front propagates from one side to the other. There is a measurable ClO$_2$ concentration only behind the front thus disinfection of the slab is completed only when the reaction front reaches the other side of the slab. The characteristic time *T* required for that can be calculated by the parabolic rate law. The result:

$$T = \frac{s_0}{2c_0 D} d^2 \quad \text{that is} \quad T \propto d^2. \quad (S12)$$

That is the characteristic time *T* is proportional with the square of the thickness for a given set of non-geometrical parameters $s_0$, $c_0$, and *D*. Here *D* is the diffusion coefficient of C in medium M. Alternatively, the penetration depth *p* of a sharp reaction is proportional with the square root of the time *t*:

$$p = \sqrt{\frac{2c_0 D}{s_0}} \cdot \sqrt{t} \quad \text{that is} \quad p \propto \sqrt{t}. \quad (S13)$$

Naturally the above formula gives the right *p* value only when $t \leq T$ or when *d* is infinitely long.

<u>Derivation of the parabolic rate law for a slab (or membrane)</u>
As the concentration profile in Fig. S1 shows, it is assumed that the reaction occurs only in the plane *x = p*. This can be a good approximation if most of the reaction takes place in a narrow reaction zone much thinner than *d* (which is valid for a fast reaction combined with a relatively slow diffusion).
The ClO$_2$ current $I_C$ across the slab with cross-section *A* in the region 0 < *x* < *p* can be given by Fick's law of diffusion:

$$I_C = -A \cdot D \cdot \frac{dc}{dx} \quad (S14)$$

($I_C$ is positive when the ClO$_2$ flow points from left to right in Fig. S1).
In a quasi steady state

$$\frac{\partial c}{\partial t} \approx 0 \quad \frac{\partial c}{\partial t} = D \frac{\partial^2 c}{\partial x^2} \quad \Rightarrow \quad \frac{\partial^2 c}{\partial x^2} \approx 0 \quad \Rightarrow \quad \frac{\partial c}{\partial x} = const, \quad (S15)$$

we can assume a linear concentration profile and so

$$I_C \approx A \cdot D \cdot \frac{c_0}{p}. \qquad (S16)$$

Next we can apply the component balance. If $N_S$ is the mole number of the remaining S molecules in the volume $V$ ($N_S = s_0 V$) then we can write

$$\frac{dN_S}{dt} = \frac{d(s_0 V)}{dt} = \frac{d(s_0 A(d-p))}{dt} = -I_C$$

$$-s_0 A \frac{dp}{dt} = -A \cdot D \cdot \frac{c_0}{p} \quad \Rightarrow \quad p \frac{dp}{dt} = D \cdot \frac{c_0}{s_0}$$

$$\int_0^d p\, dp = D \cdot \frac{c_0}{s_0} \int_0^T dt \quad \Rightarrow \quad \frac{d^2}{2} = D \cdot \frac{c_0}{s_0} \cdot T \quad \Rightarrow \qquad (S17)$$

$$\Rightarrow \quad T = \frac{s_0}{2 c_0 D} \cdot d^2$$

which is the parabolic rate law in one dimension for a slab.

<u>The parabolic rate law for an infinitely long cylinder of radius $R$</u>
In this case, the characteristic time $T$ is when the sharp reaction front starting from the surface propagating inward reaches the symmetry axis of the cylinder.

$$T = \frac{s_0}{4 c_0 D} R^2 \quad \text{that is} \quad T \propto R^2 \qquad (S18)$$

We shall regard concentration distributions with cylindrical symmetry where the local concentration $c$ is a function of the radius $r$ only – that is $c=c(r)$ – and independent of the azimuthal angle $\varphi$ and the height $z$. In an analogy to the one dimensional case

$$I_C = -A \cdot D \cdot \frac{dc}{dr} = -2 r \pi H \cdot D \cdot \frac{dc}{dr}, \qquad (S19)$$

where $H$ is the height of the cylinder. We will assume a quasi steady state concentration in the zone of $R > r > R-p$, where $p$ is the penetration depth. If $I_C$ is independent of $r$ in this region then $r \cdot \frac{dc}{dr} = const$. Regarding the boundary conditions: $c(R-p) = 0$ and $c(R) = c_0$ the steady state concentration profile in this region can be written as:

$$c = c_0 \cdot \frac{\ln\left(\dfrac{r}{R-p}\right)}{\ln\left(\dfrac{R}{R-p}\right)}, \qquad (S20)$$

and

$$\frac{dc}{dr} = \frac{1}{r} \cdot \frac{c_0}{\ln\left(\dfrac{R}{R-p}\right)}. \qquad (S21)$$

Thus, with the above quasi steady state approximation

$$I_C = -2r\pi H \cdot D \cdot \frac{dc}{dr} \approx -2\pi H \cdot D \cdot \frac{c_0}{\ln\left(\frac{R}{R-p}\right)}. \tag{S22}$$

The negative sign shows that $I_C$ points inward: it is negative when $\frac{dc}{dr} > 0$.

Next we can apply the component balance. If $N_S$ is the mole number of the remaining S molecules in the volume $V$ ($N_S = s_0 V$) then we can write

$$\frac{dN_S}{dt} = \frac{d(s_0 V)}{dt} = \frac{d\left(s_0 H (R-p)^2 \pi\right)}{dt} = -|I_C|$$

$$-s_0 H \pi 2 (R-p) \frac{dp}{dt} = -2\pi H \cdot D \cdot \frac{c_0}{\ln\left(\frac{R}{R-p}\right)} \Rightarrow (R-p)\ln\left(\frac{R}{R-p}\right)\frac{dp}{dt} = D \cdot \frac{c_0}{s_0} \tag{S23}$$

$$\int_0^R (R-p)\ln\left(\frac{R}{R-p}\right) dp = D \cdot \frac{c_0}{s_0} \int_0^T dt \Rightarrow \frac{R^2}{4} = D \cdot \frac{c_0}{s_0} \cdot T \Rightarrow$$

$$\Rightarrow T = \frac{s_0}{4 c_0 D} \cdot R^2$$

The parabolic rate law in three dimensions for a sphere of radius R

$$T = \frac{s_0}{6 c_0 D} R^2 \quad \text{that is} \quad T \propto R^2 \tag{S24}$$

We shall regard concentration distributions with spherical symmetry where the local concentration $c$ is a function of the radius $r$ only – that is $c=c(r)$ – and independent of the azimuthal angle $\varphi$ and the polar angle $\Theta$. In an analogy to the one dimensional case,

$$I_C = -A \cdot D \cdot \frac{dc}{dr} = -4r^2 \pi \cdot D \cdot \frac{dc}{dr}. \tag{S25}$$

We will assume a quasi steady state concentration in the zone of $R > r > R-p$ where $p$ is the penetration depth. If $I_C$ is independent of $r$ in this region, then $r^2 \cdot \frac{dc}{dr} = const$. Regarding the boundary conditions: $c(R-p) = 0$ and $c(R) = c_0$ the steady state concentration profile in this region can be written as:

$$c = c_0 \frac{R}{p}\left(1 - \frac{R-p}{r}\right), \tag{S26}$$

and

$$\frac{dc}{dr} = c_0 \frac{R}{p} \cdot \frac{R-p}{r^2}. \tag{S27}$$

Then with the quasi steady state approximation:

$$I_C = -4r^2\pi \cdot D \cdot \frac{dc}{dr} \approx -4\pi \cdot D \cdot c_0 \frac{R}{p} \cdot (R-p). \qquad (S28)$$

The negative sign shows again that $I_C$ points inward: it is negative when $\frac{dc}{dr} > 0$.

Next we can apply the component balance. If $N_S$ is the mole number of the remaining S molecules in the volume $V$ ($N_S = s_0 V$), then we can write

$$\frac{dN_S}{dt} = \frac{d(s_0 V)}{dt} = \frac{d\left(s_0 4(R-p)^3 \pi/3\right)}{dt} = -|I_C|$$

$$-s_0 \pi 4(R-p)^2 \frac{dp}{dt} = -4\pi \cdot D \cdot c_0 \frac{R}{p} \cdot (R-p) \quad\Rightarrow\quad \frac{p(R-p)}{R}\frac{dp}{dt} = D \cdot \frac{c_0}{s_0}$$

$$\int_0^R \frac{p(R-p)}{R} dp = D \cdot \frac{c_0}{s_0} \int_0^T dt \quad\Rightarrow\quad \frac{R^2}{6} = D \cdot \frac{c_0}{s_0} \cdot T \quad\Rightarrow \qquad (S29)$$

$$\Rightarrow\quad T = \frac{s_0}{6 c_0 D} \cdot R^2$$

### Quasi steady state solution of the RD equation in one dimension when the ClO$_2$ – substrate reaction is slow

The one dimensional reaction-diffusion equation (RDE) in a steady state is

$$0 = -r + D\frac{d^2 c}{dx^2}, \qquad (S30)$$

where $r$ is the rate of the reaction (R4)

$$C + S2 \rightarrow P2 \qquad (R4)$$

between the mobile reactant C and the fixed substrate S2 giving the product P2

$$r = k_2 \cdot c \cdot s_2. \qquad (S31)$$

In this case, however, it will be assumed that the rate of the reaction is relatively slow and the substrate S2 is in such a great excess that its consumption can be neglected during the time of the measurement. That is

$$r \approx k_2 \cdot c \cdot (s_2)_0, \qquad (S32)$$

and the steady state RDE:

$$\frac{d^2 c}{dx^2} = \frac{k_2 \cdot (s_2)_0}{D} \cdot c. \qquad (S33)$$

If we introduce the following notation

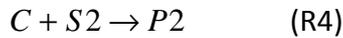

$$\frac{k_2 \cdot (s_2)_0}{D} \equiv k^2,$$

then the steady state RDE has the following form:

$$\frac{d^2 c}{dx^2} = k^2 \cdot c. \qquad (S34)$$

The general solution of the above differential equation is

$$c = c_1 \cdot e^{+k \cdot x} + c_2 \cdot e^{-k \cdot x}, \qquad (S35)$$

where the integration constants $c_1$ and $c_2$ can be calculated from the boundary conditions:

$$c(0) = c_0 \quad \text{and} \quad c(d) = 0.$$

The solution regarding the above boundary conditions:

$$c = c_0 \frac{sh(k \cdot (d - x))}{sh(k \cdot d)}. \quad (S36)$$

If the reaction is very slow, then the approximation

$$sh(\alpha) \approx \alpha \quad \text{if} \quad \alpha << 1 \quad (S37)$$

can be applied and the steady state concentration profile obtained this way is the linear concentration profile valid for pure diffusion in the absence of any chemical reaction:

$$c = c_0 \left(1 - \frac{x}{d}\right). \quad (S38)$$

The current density $j$ of component C leaving the slab is maximal when there is no chemical reaction:

$$j_{MAX} = -D \frac{dc}{dx} = D \frac{c_0}{d}. \quad (S39)$$

The current density is smaller if there is a slow reaction in the slab:

$$j_{RD} = -D \left(\frac{dc}{dx}\right)_{x=d} = D \cdot c_0 \frac{k}{sh(kd)}. \quad (S40)$$

Finally, the ratio of the current densities is:

$$\frac{j_{RD}}{j_{MAX}} = \frac{kd}{sh(kd)}. \quad (S41)$$

**Supplementary Table 1**

Data depicted in Figure 2.

$V$ is the cumulative volume of the 0.01 M $Na_2S_2O_3$ titrant added until time $t$.

| 1st exp | 2nd exp | |
|---|---|---|
| t / s | t / s | V / ml |
| 0 | 0 | 0 |
| 540 | 140 | 0.5 |
| 630 | 202 | 1 |
| 695 | 252 | 1.5 |
| 752 | 294 | 2 |
| 809 | 326 | 2.5 |
| 855 | 366 | 3 |
| 876 | 403 | 3.5 |
| 915 | 440 | 4 |
| 955 | 476 | 4.5 |
| 995 | 507 | 5 |
| 1038 | 542 | 5.5 |
| 1078 | 575 | 6 |
| 1118 | 608 | 6.5 |
| 1153 | 642 | 7 |
| 1192 | 675 | 7.5 |
| 1226 | 708 | 8 |
| 1262 | 747 | 8.5 |
| 1299 | 782 | 9 |
| 1342 | 814 | 9.5 |
| 1374 | 847 | 10 |
| 1412 | 880 | 10.5 |
| 1450 | 908 | 11 |
| 1480 | 940 | 11.5 |
| 1518 | 972 | 12 |
| 1552 | 1004 | 12.5 |
| 1588 | 1038 | 13 |

**Supplementary Table 2**

Data depicted in Figure 3.

$V$ is the cumulative volume of the 0.01 M $Na_2S_2O_3$ titrant added until time $t$.

| 1st day | 2nd day | 3rd day | |
|---|---|---|---|
| t / s | t / s | t / s | V / ml |
| 0 | 0 | 0 | 0 |
| 2490 | 530 | 250 | 0.5 |
| 2894 | 655 | 346 | 1 |
| 3198 | 750 | 418 | 1.5 |
| 3456 | 837 | 493 | 2 |
| 3670 | 916 | 568 | 2.5 |
| 3882 | 991 | 639 | 3 |
| 4080 | 1069 | 713 | 3.5 |
| 4273 | 1140 | 777 | 4 |
| 4452 | 1216 | 856 | 4.5 |
| 4636 | 1285 | 928 | 5 |
| 4812 | 1354 | 1003 | 5.5 |
| 4993 | 1420 | 1076 | 6 |
| 5171 | 1487 | 1147 | 6.5 |
| 5344 | 1556 | 1216 | 7 |
| 5522 | 1618 | 1286 | 7.5 |
| 5704 | 1680 | 1358 | 8 |
| 5889 | 1743 | 1427 | 8.5 |
| 6075 | 1810 | 1501 | 9 |
| 6266 | 1877 | 1579 | 9.5 |
| 6472 | 1942 | 1655 | 10 |
| 6658 | 2008 | 1734 | 10.5 |
| 6852 | 2077 | 1819 | 11 |
| 7059 | 2146 | 1894 | 11.5 |
| | 2218 | 1978 | 12 |
| | 2280 | 2061 | 12.5 |
| | 2346 | 2144 | 13 |
| | | 2235 | 13.5 |

**Supplementary Table 3**

Bacterial spectrum cultured from the wounds shown in Figures 4–6. Samples were taken when changing the bandage before applying $ClO_2$.

| Patient I. | at the admission | Pseudomonas aeruginosa<br>Escherichia coli<br>Enterococcus faecalis |
|---|---|---|
| | after 3 days | Proteus mirabilis |
| | after 14 days | Streptococcus alfa-hemoliticus |
| | after 28 days | Staphylococcus aureus |
| Patient II. | at the admission | Staphylococcus aureus |
| | after 1 week | no bacteria were found |
| | after 2 weeks | Corynebacterium species |
| Patient III. | at the admission | Staphylococcus aureus<br>Gram negative rod |
| | after 1 week | no pathogens were found |
| | after 4 weeks | Morganella morganii |

**Comments to Supplementary Table 3**

i) As can be seen, due to the application of $ClO_2$ the bacterial flora changes. It can even disappear for a while (as in the case of Patient II and III) but usually only the species are replaced with different ones. The rate of wound healing, however, is not affected by the presence or the absence of certain bacteria according to our observations. This is in accordance with other authors' experience who report that the presence of micro-organisms at the wound site is neither deleterious to healing nor is it indicative that infection is inevitable[7]. For instance, 69 bacterial species were isolated in a study involving 58 patients with venous leg ulcers, none of them displaying clinical signs of infection[8].

ii) James et al.[9] provided evidence that chronic wounds – unlike acute wounds – are colonised with biofilms. Microscopic analysis of debridement specimens from chronic wounds revealed the presence of densely aggregated microorganisms encapsulated in a protective matrix consistent with biofilms[9]. Most probably $ClO_2$ is promoting wound healing by killing all the bacteria on the surface including the ones living in biofilms. However, as the contact time of $ClO_2$ is short due to its volatility, it cannot prevent the reappearance of bacteria, but its periodic application prevents the formation of a new biofilm.

iii) It is worth mentioning that among the 50 patients treated with $ClO_2$ there were two cases where the wounds were colonized with MRSA and $ClO_2$ was able to decolonize and heal these wounds also.